\documentclass[10pt]{iopart}

\usepackage{graphicx}
\usepackage{amssymb}
\usepackage{exscale}
\usepackage{iopams}
\usepackage{subfigure}
\usepackage{xspace}

\newfont{\tensy}{cmsy10}
\newcommand{\chem}[1]{{$\fontdimen16\tensy=3.0pt
    \fontdimen17\tensy=3.0pt \mathrm{#1}$}}

\newcommand{\ie}[0]{i.e.\@\xspace}

\newcommand{\Tc}[0]{T_{\mathrm{C}}}

\newcommand{\up}[0]{\uparrow} 
\newcommand{\Nup}[0]{\downarrow} 
\newcommand{\en}[0]{\epsilon}
\newcommand{\ek}[0]{\epsilon_{\bi{k}}}
\newcommand{\om}[0]{\omega}
\newcommand{\Si}[0]{\Sigma}
\newcommand{\si}[0]{\sigma}
\newcommand{\las}[0]{\langle}
\newcommand{\ras}[0]{\rangle}
\newcommand{\la}[0]{\left\las}

\newcommand{\ra}[0]{\right\ras}

\newcommand{\ket}[1]{\left|#1\ra}  
\newcommand{\bra}[1]{\la#1\right|} 
\newcommand{\braket}[2]{\la#1\left.\right|#2\ra} 
\newcommand{\com}[2]{[#1,#2]}
\newcommand{\res}[0]{m$\Omega$cm\xspace}
\newcommand{\con}[0]{(m$\Omega$cm)$^{-1}$\xspace}
\renewcommand{\Im}[0]{\mathrm{Im}\ }

\begin{document}


\title[Spectral properties of the Holstein-DE model]{Spectral properties of
  the Holstein double-exchange model and application to manganites}

\author{M Hohenadler\footnote[1]{Present address: Institut f\"ur Theoretische
    Physik, Technische Universit\"at Graz, Graz A-8010, Austria} and D M
    Edwards}
\address{%
  Department of Mathematics, Imperial College, London SW7 2BZ, United
  Kingdom}

\begin{abstract}
  Calculations of one-electron spectral functions, optical conductivity and
  spin-wave energy in the Holstein double-exchange model are made using the
  many-body coherent potential approximation. Satisfactory agreement is
  obtained with angle-resolved photoemission results on
  \chem{La_{1.2}Sr_{1.8}Mn_2O_7} and optical measurements on
  \chem{Nd_{0.7}Sr_{0.3}MnO_3}. A pseudogap in the one-electron spectrum at
  the Fermi level plays an important role in both systems, but a
  small-polaron band is only predicted to exist in the La system. A rigorous
  upper bound on spin-wave energies at $T=0$ is derived. The spin-wave
  stiffness constant $D$ decreases with increasing electron-phonon coupling
  $g$ in a similar way to the Curie temperature $\Tc$, but $D/(k_{\rm B}\Tc)$
  increases for large $g$ (low $\Tc$) as observed experimentally.
\end{abstract}



\section{Introduction}\label{sec:introduction}

Recently there has been much interest in the manganite compounds
\chem{R_{1-x}A_xMnO_3}, where R is a rare-earth element such as La or Nd and
A is a divalent metal ion such as Ca, Sr or Pb. These compounds are generally
ferromagnetic for $x\simeq0.2-0.4$ and in many of them, near the Curie
temperature $\Tc$, the electrical resistivity $\rho$ decreases strongly in an
applied magnetic field. This effect is known as colossal magnetoresistance
(CMR) and a recent review is by Ramirez \cite{Ra97}. Ferromagnetism in these
compounds is believed to be due to the ``double exchange'' (DE) mechanism,
which operates when local spins are strongly coupled, by Hund's rule, to the
spins of itinerant electrons occupying a narrow band. The local spins, of
magnitude $S=3/2$, correspond to three localized Mn d electrons of \chem{t_{2g}}
symmetry and the band is derived from Mn d states of \chem{e_g} symmetry. The band
contains $n=1-x$ electrons per atom. Millis \etal \cite{MiLiSh95} stressed
that to describe the physics of the manganites completely it is necessary to
consider coupling of the electrons to local phonons as well as to local
spins. Subsequently, Millis \etal \cite{MiMuSh96II} made detailed
calculations of such a model, with local spins and local moments treated
classically, using dynamical mean field theory (DMFT). Recently Green
\cite{Gr01}, extending a many-body coherent potential approximation (CPA)
developed by Edwards \etal \cite{EdGrKu99,GrEd99} for the pure DE model,
considered a similar model with local spins and local phonons treated
quantum-mechanically. In the classical spin limit of the DE model the
many-body CPA is equivalent to the DMFT of Furukawa \cite{Fu94,Fu96}. Green
\cite{Gr01} discussed a number of physical properties of the model with
electron-phonon coupling, which he called the Holstein-DE model. These
include the resistivity, and its dependence on the applied magnetic field
(the CMR effect) and on pressure, and the Curie temperature $\Tc$. In
section~\ref{sec:many-body-cpa} we summarize the many-body CPA treatment of
the Holstein-DE model and comment on the application of Green's results to
the manganites. The aim of this paper is to calculate some spectral
properties of the model and to compare with experimental results. We restrict
our attention to the paramagnetic state and the ferromagnetic ground state
where the many-body CPA simplifies considerably. In
section~\ref{sec:angle-resolv-phot} we calculate spectral functions which
relate to the results of angle-resolved photoemission (ARPES) and in
section~\ref{sec:optical-conductivity} we consider the optical conductivity.
In section~\ref{sec:spin-wave-spectrum} we calculate the spin-wave stiffness
constant in the ferromagnetic state at $T=0$ and show how it decreases with
increasing electron-phonon coupling in a similar way to $\Tc$. A brief
summary is given in section~\ref{sec:summary}.

\section{Many-body CPA for the Holstein-DE model}\label{sec:many-body-cpa}

The Hamiltonian of the Holstein-DE model in the absence of an applied
magnetic field is
\begin{equation}\label{eq:h_hde}
    H = \sum_{ij\si} t_{ij}c_{i\si}^{\dagger}c_{j\si}-J\sum_i
    \mathbf{S}_i\cdot\bsigma_i
    -g\sum_i n_i\left(b_i^{\dagger}+b_i\right)+\om\sum_i
    b_i^{\dagger}b_i\,.
\end{equation}
The first two terms constitute the DE model and the first, third and fourth
terms form the Holstein model \cite{Ho59b}. Einstein phonons on site $i$,
with energy $\omega$ and creation operator $b_i^\dag$ couple to the electron
occupation number $n_i=\sum_\si n_{i_\si}$ with coupling strength $g$. Here
$n_{i\si}=c_{i\si}^\dag c_{i\si}$, where $c_{i\si}^\dag$ creates an electron
of spin $\si$ on lattice site $i$. The conduction electron spin
$\bsigma_i=(\si_i^x,\si_i^y,\si_i^z)$ couples to the local spin $\bi{S}_i$
with Hund exchange parameter $J>0$, and $t_{ij}$ is the band hopping
integral. The components of $\bsigma_i$ are defined as
\begin{equation}\fl
  \si_i^+=\si_i^x+i\si_i^y=c_{i\up}^\dag
  c_{i\Nup}\,,\,\si_i^-=\si_i^x-i\si_i^y=c_{i\Nup}^\dag
  c_{i\up}\,,\,\si_i^z=\frac{1}{2} \left(n_{i\up}-n_{i\Nup}\right)\,.
\end{equation}
It is reasonable to assume that the manganites are in the double exchange
limit $J\gtrsim W$ \cite{Sa96}, where $2W$ is the width of the itinerant
electron band, and in this paper we assume $J=\infty$. This ensures that
there is no double occupation of a site by electrons, so that the system is a
Mott insulator for $n=1$. This is not so for the two-band model used by
Millis \etal \cite{MiLiSh95}, and Held and Vollhardt \cite{HeVo00} have
stressed the need to introduce a strong on-site Coulomb interaction in this
case. Clearly the one-band model neglects effects of \chem{e_g} orbital
degeneracy such as a proper treatment of the Jahn-Teller effect and the
orbital ordering which occurs in the undoped system ($n=1$). However the
nature of the local phonon mode in equation~(\ref{eq:h_hde}) is not specified
and could correspond to a tetrahedral distortion of the oxygen octahedron
surrounding a Mn site as in the dynamical Jahn-Teller effect. 

An important feature of the many-body CPA for the one-particle retarded Green
function is that it becomes exact in the atomic limit $t_{ij}=0$. In this
limit, and with $J=\infty$, the Green function is given by \cite{Gr01}
\begin{eqnarray}\label{eq:G_AL}\fl\nonumber
  G^{\rm AL}(\en)=\sum_{r=-\infty}^{\infty}\frac{{\rm I}_r\left\{2\lambda
  \left[b(\om)(b(\om)+1)\right]^{1/2}\right\}}
  {(2S+1)\exp\{\lambda\left[2b(\om)+1\right]\}}\\
  \times\frac{(2S+1)\frac{n}{2}\exp\left(r\beta\om/2\right)+(S+1)(1-n)\exp\left(-r
  \beta\om/2\right)}{\en+r\om}\,,
\end{eqnarray}
where $\mathrm{I}_r$ is the modified Bessel function, $\lambda=g^2/\om^2$ and
$b(\om)=(\exp(\beta\om)-1)^{-1}$ is the Bose function with $\beta=(k_{\rm
  B}T)^{-1}$. In taking the limit $J\rightarrow\infty$ we have made a shift
of energy origin $\en\rightarrow\en-JS/2$ and the polaron binding energy
$\lambda\om$ is also absorbed in the chemical potential. The form of the
many-body CPA used by Green \cite{Gr01} for finite band-width becomes
particularly simple in the paramagnetic state. For a band density of states
of elliptic form $D_{\rm e}(\en)=[2/(\pi W^2)]\sqrt{W^2-\en^2}$, the local
Green function $G(\en)$ satisfies the CPA equation
\begin{equation}\label{eq:self_cons}
  G(\en)=G^{\rm AL}(\en-W^2 G/4)\,.
\end{equation}
Furthermore the self-energy $\Si(\en)$ is related to the local Green function
by the equation \cite{EdGrKu99}
\begin{equation}\label{eq:1}
  \Si(\en)=\en-G^{-1}-W^2 G/4\,.
\end{equation}
As well as the paramagnetic state, we shall also consider the case of a
completely saturated ferromagnetic state at zero temperature, with all local
and itinerant spins aligned. Then the double-exchange term in
equation~(\ref{eq:h_hde}) becomes merely a constant shift in energy and the
Hamiltonian is equivalent to that of the pure Holstein model. Within the
many-body CPA the saturated state is actually the self-consistent ground
state only for $S=\infty$ \cite{Gr01} and we shall mainly consider this
limit. It is found that in the DE model \cite{EdGrKu99,GrEd99} neither $\Tc$
nor the resistivity vary enormously with $S$ so that this is a reasonable
approximation to the $S=3/2$ Mn spin. In the saturated ferromagnetic state at
$T=0$, with all $n$ electrons per atom having $\up$ spin, the local Green
function $G_{\up}(\en)$ again satisfies equation~(\ref{eq:self_cons}) with
$G^{\rm AL}$ replaced by
\begin{equation}\label{eq:G_AL_T0}
  G_{\up}^{\rm AL}(\en)=\rme^{-\lambda}\left\{\frac{1}{\en}+\sum_{r=1}^{\infty}
      \frac{\lambda^r}{r!}\left(\frac{n}{\en+\om r}+\frac{1-n}{\en-\om
      r}\right)\right\}\,.
\end{equation}

The many body CPA is successful in describing the crossover from weak
electron-phonon coupling, through intermediate coupling where small-polaron
bands begin to appear, to strong coupling where some results similar to those
of standard small-polaron theory are recovered. It therefore extends the work
of Millis \etal \cite{MiMuSh96II}, where phonons are treated classically, to
the quantum small-polaron regime. One important quantum effect on a
thermodynamic property is the behaviour of $\Tc$ for strong electron-phonon
coupling. Millis \etal \cite{MiMuSh96II} find $\Tc\sim g^{-4}$ whereas Green
\cite{Gr01} finds $\Tc$ is exponentially small for $g/W\gtrsim0.35$. The
physics is dominated by a very narrow polaron band. Green \cite{Gr01} showed
how the Holstein-DE model could describe the very different behaviour of
\chem{La_{1-x}Sr_xMnO_3} (LSMO) and \chem{La_{1-x}Ca_xMnO_3} (LCMO), with
$x\sim0.33$ where $\Tc$ is largest. In LSMO the resistivity $\rho$ increases
monotonically with temperature $T$, with a metal-poor metal transition at
$\Tc$, whereas LCMO shows a metal-insulator transition with $\rho$ decreasing
with $T$ above $\Tc$. Also $\rho(\Tc)$ is an order of magnitude smaller in
LSMO than in LCMO, This type of behaviour can be understood in the
Holstein-DE model by assuming that $g/W$ is about $50\%$ larger in LCMO than
in LSMO and, within the model, this is consistent with the considerably lower
$\Tc$ in LCMO. (The form of $\rho(T)$ in figure~7 of reference~4 is
indistinguishable for $g/W=0.1$ from that shown for $g/W=0.01$ \cite{Green}
and contrasts strongly with the curve in figure~6 for $g/W=0.16$). The
calculated resistivity is not as sensitive to applied magnetic field and
pressure as observed but this discrepancy might be removed by introducing a
decrease in $g/W$ with decreasing resistivity owing to enhanced screening of
the ionic charges \cite{Gr01}. A defect of the CPA treatment is its failure
to describe coherent Bloch-like states in the saturated ferromagnetic state
at $T=0$, leading to a spurious residual resistivity \cite{Gr01}.

In this paper we calculate further properties of the Holstein-DE model and
confront them with experiment.

\section{Angle-resolved photoemission spectroscopy (ARPES)}
\label{sec:angle-resolv-phot}

The decreasing resistivity of LCMO as the temperature increases above $\Tc$
is, according to Millis \etal \cite{MiMuSh96II} and Green \cite{Gr01}, due to
the gradual filling of a pseudogap in the density of states. In Green's work
the pseudogap contains well-defined polaron sub-bands in the (hypothetical)
paramagnetic state at $T=0$ but above $\Tc$, with parameters appropriate to
LCMO, these are smeared out so as to resemble the classical picture of Millis
\etal. The pseudogap should be observable in ARPES measurements and in
optical conductivity. In an excellent paper on ARPES for the bilayer
manganite \chem{La_{1.2}Sr_{1.8}Mn_2O_7}, nominally with $n=0.6$, Dessau
\etal \cite{De98} interpret their results very much in the spirit of the
Holstein model. In this section we present some calculations of the spectral
functions in the low temperature ferromagnetic state to compare with this
experimental data. For convenience we take $n=0.5$ and also $S=J=\infty$, as
discussed above. The charge ordering which might occur for $n=0.5$ is suppressed since our 
treatment imposes spatial homogeneity. With this constraint there is no
qualitative difference between systems with $n=0.5$ and with $n=0.3$ or $0.6$, for
example. We take the phonon energy to be $\om/W=0.05$, the same
typical value used in previous calculations \cite{Gr01}. A half-width $W=1$
eV is consistent with the \chem{e_g} bands crossing the Fermi level in the
calculations of Dessau \etal \cite{De98}, shown in
figure~\ref{fig:weights:b}. The low $\Tc=126$ K in this bilayer manganite is
partly due to quasi-two dimensional fluctuations, but the large resistivity
$\rho\simeq3$ \res at low temperatures indicates that small-polaron bands
might exist even in the ferromagnetic state. Consequently the electron-phonon
coupling should be stronger than in cubic manganites like LCMO and we choose
$g/W=0.2$.

\begin{figure}[htbp]
  \centering \includegraphics[width=0.5\textwidth]{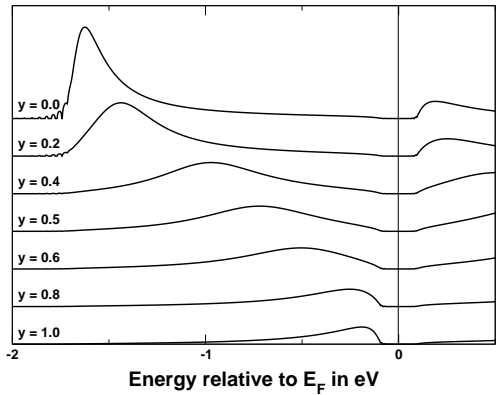}
\caption{\label{fig:spec_func_th}%
  The spectral function $A_{\bi{k}}(\en)$ in the ferromagnetic state at $T=0$
  for $J=S=\infty$, $n=0.5$ and strong electron-phonon coupling $g/W=0.20$,
  with $\bi{k}=\pi(1,y)$.}
\end{figure}

\begin{figure}[htbp]
  \centering \includegraphics[width=0.5\textwidth]{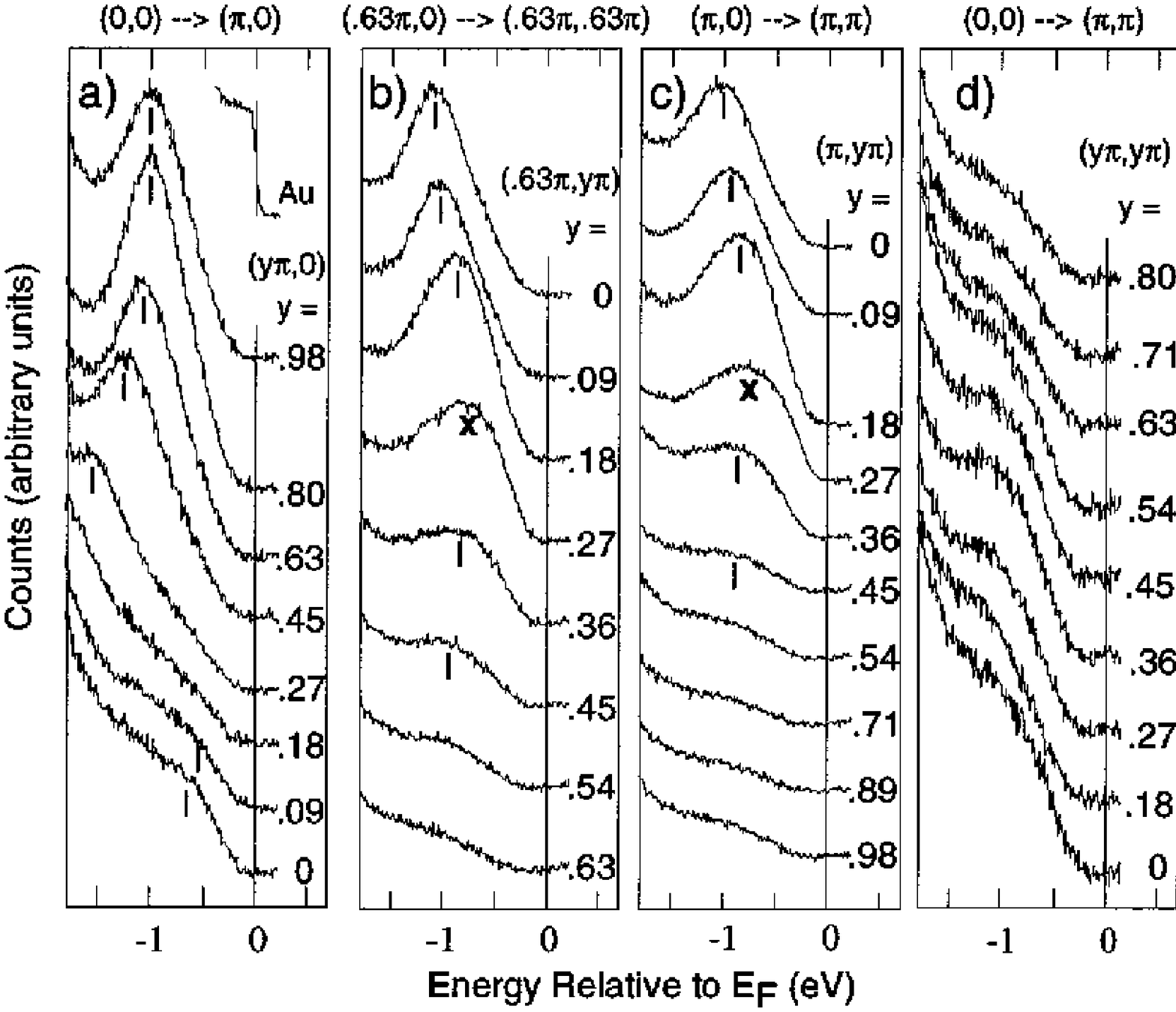}
\caption{\label{fig:spec_func_exp}%
  ARPES spectra of \chem{La_{1.2}Sr_{1.8}Mn_{2}O_{7}} ($\Tc=126\,{\rm K}$) in
  the ferromagnetic state at $T=10\,{\rm K}$, reproduced from Dessau \etal
  \cite{De98}.}
\end{figure}

\begin{figure}[htbp]
  \centering \subfigure[]{
    \label{fig:weights:a}
    \includegraphics[width=0.4\textwidth]{weights.eps}} \subfigure[]{
    \label{fig:weights:b}
    \includegraphics[width=0.4\textwidth]{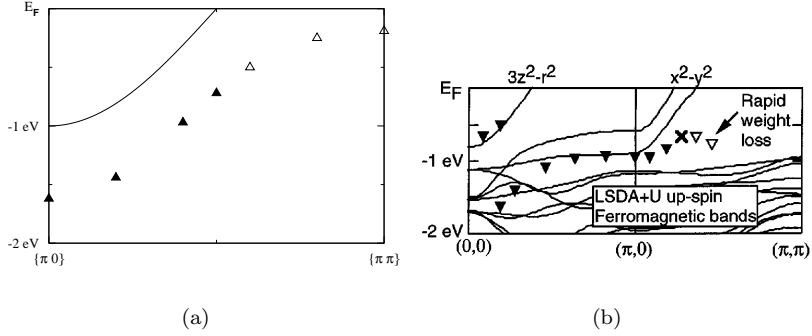}}
  \caption{\label{fig:weights}%
    (a) the band energy $\ek$ (\full) and the position of the peak centres in
    figure~\ref{fig:spec_func_th}. Different symbols denote high
    ($\fulltriangle$) and low ($\opentriangle$) spectral weight, obtained by
    integration over the spectral function up to the Fermi energy $E_{\rm F}$. (b)
    virtual crystal LSDA+U majority spin bands for
    \chem{La_{1.2}Sr_{1.8}Mn_2O_7} with experimental peaks from
    figure~\ref{fig:spec_func_exp}, reproduced from \cite{De98}.}
\end{figure}

The one-electron spectral function is given by
\begin{eqnarray}\label{eq:spec_func}\nonumber
A_{\bi{k}}(\en)&=-\pi^{-1}\Im [\en-\ek-\Si(\en)]^{-1}\\
  &=-\pi^{-1}\Si_{\en}''/[(\en-\ek-\Si'_\en)^2+{\Si_\en''}^2]
\end{eqnarray}
where $\Si'_\en$, $\Si''_\en$ are the real and imaginary parts of the
self-energy $\Si(\en)$ and $\ek$ is the band energy for wave-vector $\bi{k}$.
In the ferromagnetic state at $T=0$ the local Green function $G_\up(\en)$ is
calculated from equation~(\ref{eq:self_cons}), with $G^{\rm AL}$ defined by
(\ref{eq:G_AL_T0}), and $\Si(\en)$ follows from equation~(\ref{eq:1}). These
equations assume an elliptic density of states which we here regard as an
approximation to the density of states for a band which takes the form
$\ek=-W\cos\pi y$ for $\bi{k}=\pi(1,y)$, $0\le y\le1$. This band is shown as
a full line in figure~\ref{fig:weights:a} and crosses the Fermi level $E_{\rm
  F}$ at $\bi{k}=\pi(1,\frac{1}{2})$. It roughly models one of the $x^2-y^2$
bands in figure~\ref{fig:weights:b}. The calculated results for $A_\bi{k}$
are shown in figure~\ref{fig:spec_func_th}. Well away from the Fermi level, a
well-defined peak exists which broadens as $\bi{k}$ approaches the Fermi
momentum at $y=0.5$. For larger $y$ the weight below the Fermi level is
strongly reduced. The peaks never approach the Fermi level closely which is
an important feature of the observed spectra \cite{De98} reproduced in
figure~\ref{fig:spec_func_exp}. The theoretical curves in
figure~\ref{fig:spec_func_th} resemble quite closely the data of
figure~\ref{fig:spec_func_exp}(c). There is a pseudogap in the calculated
spectra extending about 0.1 eV on each side of the Fermi level. In fact this
pseudogap contains polaron bands like those shown in figure~4 of Green's
paper \cite{Gr01}. However, their amplitude is too small to show up in
figure~\ref{fig:spec_func_th} and in the experimental data. Nevertheless, it
is the central polaron band around the Fermi level which is responsible for
the low but finite conductivity of the system. The positions of the peaks in
figure~\ref{fig:spec_func_th} are plotted in figure~\ref{fig:weights:a} and
comparison can be made with the right half of figure~\ref{fig:weights:b}
reproduced from Dessau \etal \cite{De98}. Filled and unfilled symbols
correspond to high and low weights, respectively, obtained by integration of
the spectral function up to the Fermi energy. This comparison between theory
and experiment supports the conclusion of Dessau \etal \cite{De98} that, in
the manganites with a layered structure, strong electron-phonon coupling
(with the appearance of a pseudogap) is already important below $\Tc$. This
contrasts with the usual pseudocubic manganites where the pseudogap only
appears above $\Tc$. Previous work related to ours is the calculation by
Perebeinos and Allen \cite{PeAl00} of ARPES spectra in a two-band model of
undoped \chem{LaMnO_3}. It should be mentioned that Moreo \etal \cite{MoYu99}
interpret the observed pseudogap not as an intrinsic property but in terms of
phase separation.

\section{Optical conductivity}\label{sec:optical-conductivity}

In the local approximation of CPA or DMFT there is no vertex correction in
the current-current response function which may thus be expressed in terms of
the one-particle spectral function. In the paramagnetic state the optical
conductivity is given by \cite{pruschke_review,EdGrKu99}
\begin{equation}\label{eq:sigma_1}
  \si(\nu)=\frac{2\pi e^2}{3Na^3\hbar}\sum_{\bi{k}}\bi{v}_{\bi{k}}^2
  \int\,\rmd\en A_{\bi{k}\si}(\en) A_{\bi{k}\si}(\en+\nu)
  \frac{f(\en)-f(\en+\nu)}{\nu}\,,
\end{equation}
where the electron velocity $\bi{v}_\bi{k}=\nabla\ek$, $a^3$ is the volume of
the unit cell, $f(\en)=\{\exp[\beta(\en-\mu)]+1\}^{-1}$ is the Fermi function
and $N$ is the number of lattice sites. Since $A_\bi{k}$ depends on $\bi{k}$
only through $\ek$ we may define a function $\phi(\ek)$ such that
$\phi'(\ek)=A_\bi{k}(\en)A_\bi{k}(\en+\nu)$, where the dependence of $\phi$
on $\en$ and $\nu$ has been suppressed in the notation. Hence the sum in
equation~(\ref{eq:sigma_1}) may be written as
\begin{equation}\label{eq:2}
  \sum_\bi{k}\nabla\ek\cdot\nabla\phi(\ek)=-\sum_\bi{k}\phi(\ek)\nabla^2\ek\,,
\end{equation}
the last step following by Gauss's theorem. For a simple cubic tight binding
band $\ek=-2t\sum_\alpha \cos k_\alpha a$, with $\alpha$ summed over $x,y,z$,
$\nabla^2\ek=-a^2\ek$. Then the summand in equation~(\ref{eq:2}) is a
function of $\ek$ only and equation~(\ref{eq:sigma_1}) becomes
\begin{equation}\label{eq:3}
  \si(\nu)=\frac{2\pi e^2}{3a\hbar} \int\,\rmd\en \int\,\rmd E
  D_{\rm c}(E)E\phi(E)
  \frac{f(\en)-f(\en+\nu)}{\nu}
\end{equation}
where $D_{\rm c}(\en)$ is the density of states for the simple cubic band. By
using equation~(\ref{eq:spec_func}) in the definition of $\phi'(\ek)$, and
integrating with respect to $\ek$, we obtain
\begin{eqnarray}\label{eq:phi}\fl\nonumber
  \phi = \frac{1}{\pi^2} \frac{1}{ {\left( {\left(\Si''_\en -\Si''_{\en+\nu}
          \right) }^2 + O^2 \right) \left( {\left(\Si''_\en +
            \Si''_{\en+\nu} \right) }^2 +O^2\right)}}\\\nonumber
  \times\biggl\{ \Si''_{\en+\nu}\left({\Si''_{\en+\nu}}^2 - {\Si''_{\en}}^2
  {}+O^2 \right)\arctan P +\\\Si''_\en \biggl[\left( {\Si''_\en}^2
  {}-{\Si''_{\en+\nu}}^2+O^2\right) \arctan Q +\Si''_{\en+\nu}
  O \log R\biggr]\biggr\}\,,
\end{eqnarray}
with
\begin{eqnarray}\nonumber
  &O=\Si'_\en - \Si'_{\en+\nu} + \nu\,,\qquad
  &Q=\frac{\Si'_{\en+\nu} -\nu-\en+\ek}{\Si''_{\en+\nu}}\,,\\
  &P=\frac{\Si'_\en-\en+\ek}{\Si''_\en}\,,\qquad
  &R=\frac{{\Si''_\en}^2+{\left(\Si'_\en-\en+\ek\right)}^2}{{\Si''_
      {\en+\nu}}^2+{\left(\Si'_{\en+\nu}-\nu-\en+\ek \right)}^2}\,.
\end{eqnarray}
Since we calculate the self-energy $\Si$ using the elliptic density of states
$D_{\rm e}(\en)$, as discussed in section~\ref{sec:many-body-cpa}, it is
reasonable to approximate $D_{\rm c}(E)$ in equation~(\ref{eq:3}) by $D_{\rm
  e}(E)$. The integral over $E$ can then be carried out by parts so that,
using the definition of $\phi'$, we find
\begin{equation}\label{eq:4}\fl
\si(\nu)=\frac{2\pi e^2}{3 a\hbar}\int\,\rmd\en\int\,\rmd
E\frac{W^2-E^2}{3}D_{\rm e}(E)
A_E(\en)A_E(\en+\nu)\frac{f(\en)-f(\en+\nu)}{\nu}
\end{equation}
with $A_E(\en)$ given by equation~(\ref{eq:spec_func}), $\ek$ being replaced
by $E$. This is of the form given by Chung and Freericks \cite{ChFr98} and
Chattopadhyay \etal \cite{ChMiDa00}. If $D_{\rm c}(E)$ in
equation~(\ref{eq:3}) is replaced by a Gaussian, corresponding to a
hypercubic lattice in infinite dimensions, the factor $(W^2-E^2)/3$ in
equation~(\ref{eq:4}) is replaced by a constant \cite{PrCoJa93,EdGrKu99}. In
this paper we use equation~(\ref{eq:4}) which is consistent with our previous
calculations of $\si(0)$ \cite{EdGrKu99,Gr01}. This expression satisfies the
correct one-band sum rule \cite{ChMiDa00,QuCeSi98} that
$(2/\pi)\int_0^\infty\si(\nu)\rmd\nu=-Ke^2/(3a\hbar)$, where the ``kinetic
energy'' $K$ is the thermal average per lattice site of the first term in the
Hamiltonian (\ref{eq:h_hde}).

As in section~\ref{sec:angle-resolv-phot} we take $J=S=\infty$, $W=1$ eV,
$\om=50$ meV and, as previously \cite{Gr01,EdGrKu99}, $a=5${\AA}, which is
slightly larger than the Mn--Mn distance in perovskite manganites. It is
again convenient to take $n=0.5$ so that for $J=S=\infty$ the chemical
potential $\mu$ is fixed at the centre of the occupied band for all
temperatures, by symmetry. In the ferromagnetic state at $T=0$ the
spin-degeneracy factor 2 in equation~(\ref{eq:4}) is omitted.

There have been numerous optical investigations of the manganites and the
experimental data and their interpretation are varied. For photon energy
$\nu>3$ eV various peaks in the optical conductivity have been assigned to
interband transitions between the Hund's rule split bands and to
charge-transfer transitions between the O 2p and the Mn \chem{e_g} bands. The
latter do not appear in our one-band model and our assumption of $J=\infty$
eliminates the upper Hund's rule band. Our calculated $\si(\nu)$ is therefore
only non-zero in the region $\nu<2.5$ eV and in general exhibits one peak. We
shall focus the discussion by considering one material,
\chem{Nd_{0.7}Sr_{0.3}MnO_3} (NSMO), which has been investigated by at least
two experimental groups. In the paramagnetic state above $\Tc$ there is one
feature common to data on both thin films \cite{Ka96,QuCeSi98} and single
crystals \cite{LeJuLe99}. This is a broad peak at about 1.2 eV with a maximum
conductivity $\si_{\rm max}\approx0.7-0.9$ \con. However in NSMO films $\Tc$,
as deduced from the maximum in the resistivity $\rho(T)$, is larger in
oxygen-annealed films ($\Tc\approx230$ K) than in an unannealed sample
($\Tc\approx180$ K) \cite{QuCeSi98,Ka96}. In a NSMO single crystal, on the
other hand, $\rho(T)$ is essentially the same ($\Tc\approx200$ K) in polished
and annealed samples but the peak in the polished sample is shifted to
$\nu\simeq1$ eV with a reduced $\si_{\rm max}\approx0.3$ \con. Since
$\rho(T)$ and $\si(\nu)$ are quantitatively quite similar in NSMO and LCMO
\cite{QuCeSi98} we model NSMO with electron-phonon coupling $g/W=0.16$, the
same value as proposed by Green \cite{Gr01} for LCMO. The dc conductivity
$\si(0)$ at $T=10$ K is also very similar in annealed NSMO and LCMO films,
2.9 and 3.3 \con respectively \cite{QuCeSi98}. The value for NSMO agrees well
with a measured $\si(0)$ of 3.2 \con at $T=15$ K in a NSMO single crystal
\cite{LeJuLe99}.

\begin{figure}[htbp]
  \centering \subfigure[ ]{
    \label{fig:oc:a}
    \includegraphics[width=0.4\textwidth]{oc_0.16_bw.eps}} \subfigure[ ]{
    \label{fig:oc:b}
    \includegraphics[width=0.4\textwidth]{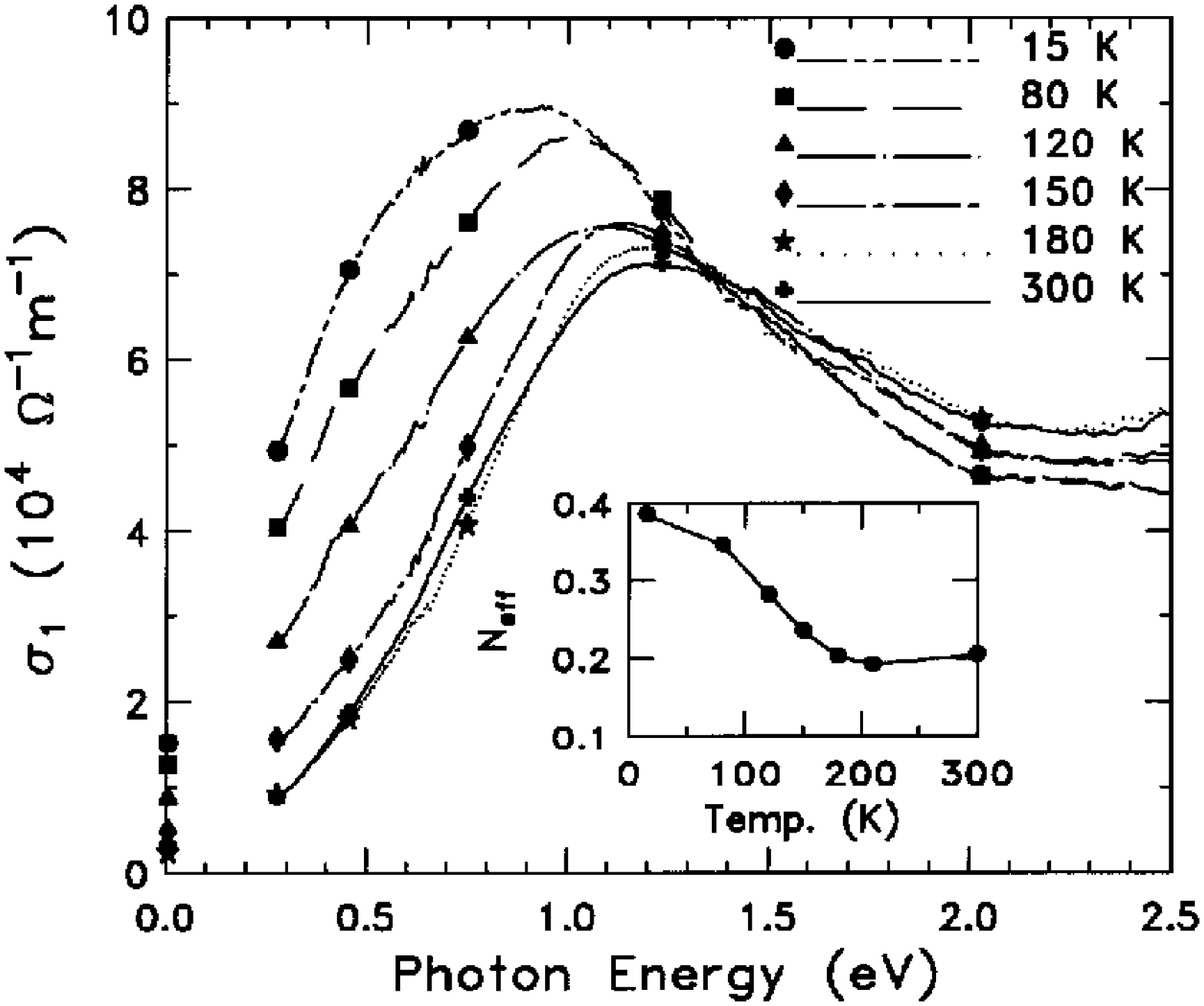}}
  \caption{\label{fig:oc}%
    (a) Calculated optical conductivity for strong electron-phonon coupling
    $g/W=0.16$ in the ferromagnetic state at $T=0$ (\full), the paramagnetic
    state at $T=\Tc$ (\dotted) and the paramagnetic state at $T=1.5\Tc$
    (\broken). The plot is for $J=S=\infty$, $n=0.5$ and $a=5${\AA}. (b)
    optical conductivity of \chem{Nd_{0.7}Sr_{0.3}MnO_3} at different
    temperatures, reproduced from Kaplan \etal \cite{Ka96}.}
\end{figure}

Using the parameters discussed above the Curie temperature of the model is
about 230 K \cite{Gr01} and in figure~\ref{fig:oc:a} we plot the calculated
optical conductivity for the ferromagnetic state at $T=0$ and the
paramagnetic state at $T=\Tc$ and $T=1.5\Tc$. The low dc conductivity of 0.11
\con at $T=0$ is due to the inadequate treatment of the coherent ground state
in the CPA. Fortuitously the incoherent scattering introduced by the CPA
seems to model quite well the low temperature incoherent scattering in the
unannealed NSMO sample of Kaplan \etal \cite{Ka96} which has a dc
conductivity of about 0.15 \con at $T=15$ K. This optical data is reproduced
in figure~\ref{fig:oc:b} for comparison with the calculated results of figure
4a. In annealed NSMO films \cite{QuCeSi98}, and in single crystals
\cite{LeJuLe99}, $\si(\nu)$ in the low temperature ferromagnetic state
continues to rise with decreasing $\nu$ down to much lower photon energy, and
$\si(0)\approx3$ \con. The above-mentioned defect of the present CPA
treatment is associated with the absence of the sharp quasi-particle peak in
the spectral function whith should exist in the ferromagnetic state at
$T=0$. Consequently the Drude peak in $\si(\nu)$ at low frequency is also absent.

As pointed out above, $\si(\nu)$ is much less sample-dependent in the
paramagnetic state above $\Tc$ and a quantitative comparison with theory is
meaningful. In figure~\ref{fig:oc:a} $\si_{\rm max}\approx0.15$ \con and it
must be remembered that this curve is the contribution of optical transitions
within the \chem{e_g} band only. To compare with the data of figure~\ref{fig:oc:b}
one should subtract the background due to other transitions which might leave
an effective $\si_{\rm max}\approx0.3$ \con. Bearing in mind the simplicity
of the one-band model, this order-of-magnitude agreement between theory and
experiment is satisfactory. Our calculated results are quite similar to those
of figure 7d in Millis \etal \cite{MiMuSh96II}. Clearly, for the present
intermediate electron-phonon coupling strength, their classical treatment of
phonons is sufficient to obtain the essential features of the optical
conductivity. The calculated peak in $\si(\nu)$ with $\nu\approx1$ eV arises
from $\bi{k}$-conserving transitions across the type of pseudogap discussed
in section~\ref{sec:angle-resolv-phot}. During the process an electron moves
from one site to a neighbouring one which was previously unoccupied. The
electron motion is accompanied by a lattice distortion, of Jahn-Teller type,
which corresponds to a displacement of the local phonon oscillator coordinate
in the Holstein-DE model. When an electron enters (leaves) a site the final
displaced (undisplaced) oscillator is generally in an excited state with
typical excitation energy $g^2/\om$. This is the atomic-limit polaron binding
energy and for the parameters assumed here is about 0.5 eV. Thus the peak in
$\si(\nu)$ occurs at about twice the polaron binding energy just as in the
standard small-polaron theory \cite{Ma90}. However for the present
intermediate electron-phonon coupling $g/W=0.16$ polaron bands near the Fermi
level are largely washed out above $\Tc$ \cite{Gr01}, so standard
small-polaron theory is not expected to apply to $\si(\nu)$ for low photon
energies. In fact an activation energy in the dc conductivity of 0.25 eV,
half the polaron binding energy as predicted by small-polaron theory
\cite{Ma90}, is about a factor 4 larger than one deduced from Green's
\cite{Gr01} numerical calculations. Green's calculation of the dc resistivity
above $\Tc$ is in reasonable agreement with experiments on NSMO where an
activation energy of about 0.08 eV is found \cite{ZhKacond-mat}. Lee \etal
\cite{LeJuLe99} quote an activation energy of about 0.15 eV and, with undue
reliance on small-polaron theory, expect to have a peak in $\si(\nu)$ at
about 0.6 eV. Although there is no sign of such a peak in their data they
claim that their one broad peak near 1.2 eV should be interpreted in terms of
a two-peak structure, one near 1.5 eV and the other below 1 eV. Our
interpretation of the 1.2 eV peak in $\si(\nu)$ for $T>\Tc$ is broadly in
line with that of several other previous authors
\cite{MiMuSh96II,QuCeSi98,Ka96}. However we stress again that, for moderate
electron-phonon coupling standard small-polaron theory does not hold at low
photon energy, where only states around the Fermi level are involved, so no
simple link between peak position and the activation energy of the dc
conductivity can be made. Below $\Tc$ the peak in $\si(\nu)$ shifts to lower
frequency as the pseudogap rapidly fills in. However in our calculations this
shift is held up due to spurious incoherent scattering in the ground state,
which limits the low-temperature dc conductivity. The same effect actually
occurs in unannealed NSMO films (figure~\ref{fig:oc}) but in annealed films
and single crystals the peak shifts almost to zero frequency
\cite{QuCeSi98,LeJuLe99}. The theoretical situation could be improved by
introducing screening effects in the electron-phonon interaction so that
$g/W$ decreases with decreasing resistivity, as mentioned at the end of
section~\ref{sec:many-body-cpa}. The spurious residual resistivity at $T=0$ drops
from 9 \res to 0.5 \res as $g/W$ decreases from 0.16 to 0.10
\cite{Gr01,Green}. Ishihara \etal \cite{IsYaNa97} and Mack and Horsch
\cite{MaHoPom} have given an alternative interpretation of the broad low
energy peak at $T=0$ in terms of orbital degrees of freedom in the
doubly-degenerate \chem{e_g} band. They propose that strong correlation, with the
constraint of no doubly occupied sites, leads to incoherent motion of the
carriers.

\section{The spin-wave spectrum}\label{sec:spin-wave-spectrum}

Quijada \etal \cite{QuCeSi98} considered the spin-wave stiffness constant $D$
in the saturated ferromagnetic state at $T=0$ and its relation to optical
conductivity. However their derivation of an expression for $D$ was
restricted to the DE model. Here we derive a simple approximate formula for
the spin-wave spectrum of the Holstein-DE model. Our main aim is to calculate
$D$ as a function of electron-phonon coupling strength and to compare with
the corresponding behaviour of $\Tc$. In the limit of $J\rightarrow\infty$
the local and itinerant spins are locked together as we make the ansatz
\begin{equation}\label{eq:wave_function}
  \ket{q}=\left(S_\bi{q}^-+\si_\bi{q}^-\right)\ket{F}\,
\end{equation}
for the state (unnormalized as yet) with a magnon of wave-vector $\bi{q}$
excited. Here $\ket{F}$ is the exact ferromagnetic ground state, which we
assume to be one of complete spin alignment, and the spin lowering operators
are defined by
\begin{equation}
  S_\bi{q}^-=\sum_i \rme^{\rmi\bi{q}\cdot\bi{R}_i}S_i^-\,,\quad
  \si_\bi{q}^-=\sum_i \rme^{\rmi\bi{q}\cdot\bi{R}_i}\si_i^-\,,
\end{equation}
where $\bi{R}_i$ is the position of lattice site $i$.

Actually we do not know the state $\ket{F}$ exactly for the Holstein-DE
model, unlike the DE model, but we shall only approximate it at a later
stage. The spin-wave energy is given by
\begin{equation}\label{eq:sw_energy}
  \om_\bi{q}=\frac{\bra{q}H\ket{q}}{\braket{q}{q}}-E_0
\end{equation}
where $E_0$ is the exact ground state energy, so that $H\ket{F}=E_0\ket{F}$.
By the variational principle, this is an upper bound on $\om_\bi{q}$. Using
equation~(\ref{eq:wave_function}) it is easy to show that
\begin{equation}\label{eq:33}
  \om_\bi{q}=\frac{\bra{q}\com{H}{S_\bi{q}^-+\si_\bi{q}^-}\ket{F}}
  {\braket{q}{q}}
\end{equation}
and $\braket{q}{q}=N(2S+n)$, where $S$ is the magnitude of the localized spin
$\bi{S}_i$. Only the first term in the Hamiltonian (\ref{eq:h_hde}) makes a
non-zero contribution to the commutator. This one-electron term may be
written as $\sum_{\bi{k}\si}\ek n_{\bi{k}\si}$ where $n_{\bi{k}\si}$ is the
occupation number for the Bloch state $\bi{k}\si$. A straight-forward
calculation yields
\begin{equation}\label{eq:disp_heisenberg}
  \om_{\bi{q}}=\frac{1}{N(n+2S)}\sum_{\bi{k}}\left(\en_{\bi{k}+\bi{q}}-
  \ek\right)\las n_{\bi{k}\up}\ras
\end{equation}
where $\las n_{\bi{k}\up}\ras=\las F|n_{\bi{k}\up}\ket{F}$. Assuming a simple
cubic tight-binding band $\ek=-t\sum_\bi{R}\rme^{\rmi\bi{k}\cdot\bi{R}}$, the
sum being over 6 nearest neighbour sites, we find
\begin{equation}\label{eq:omega_step1}
  \om_{\bi{q}}=-\frac{t}{N(n+2S)}\sum_{\bi{R}}\left(\rme^{\rmi\bi{q}
  \cdot\bi{R}}-1\right)\sum_{\bi{k}}\rme^{\rmi\bi{k}\cdot\bi{R}}\las
  n_{\bi{k}\up}\ras\,.
\end{equation}
Since $\las n_{\bi{k}\up}\ras$ has cubic symmetry in $\bi{k}$-space, the
$\bi{k}$ sum in equation~(\ref{eq:omega_step1}) is independent of the
particular neighbour $\bi{R}$. By including the factor $-t$ it may therefore
be written as $(1/6)\sum_\bi{k}\ek\las n_{\bi{k}\up}\ras$. Hence
\begin{eqnarray}\label{eq:5}\nonumber
  \om_\bi{q}&=\frac{D}{a^2}\sum_\bi{R}\left(1-\rme^{\rmi\bi{q}
  \cdot\bi{R}}\right)\\&=(2D/a^2)\left(3-\cos q_x a-\cos q_y a-\cos q_z
  a\right)\,,
\end{eqnarray}
where $\bi{q}=(q_x,q_y,q_z)$ and
\begin{equation}\label{eq:6}
  D=-\frac{a^2}{6N(n+2S)}\sum_\bi{k}\ek\las
  n_{\bi{k}\up}\ras=-\frac{Ka^2}{6(n+2S)}\,.
\end{equation}
Here $K$ is the expectation value of the kinetic energy which appears in the
optical sum rule mentioned in section~\ref{sec:optical-conductivity}. By
expanding equation~(\ref{eq:5}) in powers of $\bi{q}$ to second order, we
find $\om_\bi{q}=Dq^2$ as that $D$ defined by equation~(\ref{eq:6}) is the
spin-wave stiffness constant.

At this stage, with the expectation value $\las n_{\bi{k}\up}\ras$ calculated
in the exact ground state, equation~(\ref{eq:5}) is a rigorous upper bound on
the magnon energy for arbitrary $J$ and $S$. This is no longer the case when
we proceed to evaluate it within the many-body CPA. Clearly,
equations~(\ref{eq:5}) and~(\ref{eq:6}) apply equally to the Holstein-DE and
the DE model. However, we shall show that electron-phonon coupling in the
former model has a strong influence via $\las n_{\bi{k}\up}\ras$. In the
limit $J\rightarrow\infty$, equation~(\ref{eq:6}) is equivalent to results
derived by Nagaev \cite{Na70,Na98}, Kubo and Ohata \cite{KuOh72}, Furukawa
\cite{Fu96,FuPom}, Wang \cite{Wa98} and Quijada \etal \cite{QuCeSi98} for the
DE model. The $K$ defined in the latter paper is one-third of our kinetic
energy. Furukawa also derived the dispersion relation of
equation~(\ref{eq:5}) for the DE model. Now
\begin{equation}\label{eq:7}
  \las n_{\bi{k}\up}\ras=\int_{-\infty}^{\mu}\,\rmd\en A_\bi{k}(\en)
\end{equation}
where $A_\bi{k}(\en)$, given by equation~(\ref{eq:spec_func}), is calculated
for the saturated ferromagnetic state at $T=0$. As in deriving $\si(\nu)$ we
can replace the $\bi{k}$ sum in equation~(\ref{eq:6}) by an energy integral.
Hence, using the notation $A_E(\en)$ introduced in equation~(\ref{eq:4}),
\begin{equation}\label{eq:8}
  D=-\frac{a^2}{6(n+2S)}\int_{-W}^{W}\,\rmd E\int_{-\infty}^\mu\,\rmd\en E
  D_{\rm e}(E) A_E(\en)\,.
\end{equation}
As in the previous section we have approximated the simple cubic density of
states by the elliptic density of states $D_{\rm e}(E)$. To match the other
calculations in this paper, in particular that of the optical conductivity,
we calculate $A_E(\en)$, and hence the double integral in
equation~(\ref{eq:8}) which represents the average kinetic energy $K$, in the
limit $S=\infty$. However, in the prefactor in equation~(\ref{eq:8}) we put
$S=3/2$ as is appropriate for the localized Mn spins. This corresponds to
calculating the Bloch wall stiffness constant ($\propto D(n+2S)$), which is a
static quantity, in the limit $S\rightarrow\infty$ but retaining the
essential finiteness of the spin in the dynamical quantity $D$. Otherwise we
use the same parameters as in previous sections, except that we now take
$a=4$ {\AA}. In section~\ref{sec:optical-conductivity} we used $a=5$ {\AA} ,
for consistency with our earlier work on conductivity, but 4 {\AA} is closer
to the Mn--Mn distance in the pseudocubic manganites.

\begin{figure}[htbp]
  \centering \includegraphics[width=0.5\textwidth]{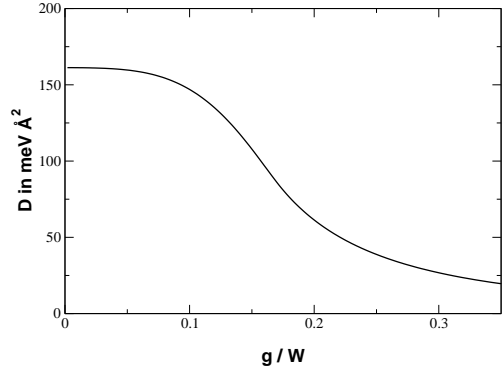}
\caption{\label{fig:spin_stiffness}%
  The spin-wave stiffness $D$ versus electron-phonon coupling $g$ in the
  saturated ferromagnetic state at $T=0$. The plot is for $S=J=\infty$, $W=1$
  eV, $n=0.5$ and $a=4$ {\AA}.}
\end{figure}

In figure~\ref{fig:spin_stiffness} we plot the spin-wave stiffness $D$ at
$T=0$ as a function of electron-phonon coupling $g/W$. The reason for the
striking decrease of $D$ with $g/W$, particularly in the range $0.1<g/W<0.2$
applicable to the manganites, is clear from equation~(\ref{eq:6}). For
$g/W=0$, the pure DE model, $\las n_{\bi{k}\up}\ras=1$ for $\bi{k}$ within
the Fermi surface and $\las n_{\bi{k}\up}\ras=0$ otherwise. The negative
quantity $K$ is the full non-interacting one-electron energy of the
ferromagnetic state which drives the double-exchange mechanism. For larger
$g/W$, $\las n_{\bi{k}\up}\ras$ is more spread out over the whole zone and
$|K|$ decreases. In an extreme limit where electrons are localized at sites,
$\las n_{\bi{k}\up}\ras$ is constant throughout the zone and hence $D=0$.
This behaviour of $D$ in the Holstein-DE model is very similar to that of
$\Tc$, as calculated by Green \cite{Gr01}. The main difference is in the
extreme strong-coupling limit where $\Tc$ becomes very small at
$g/W\approx0.35$ whereas $D$ is decreasing quite slowly. The slow decrease of
$D$ is exactly what one expects from equation~(\ref{eq:6}) and small-polaron
theory, where the kinetic energy $K\sim g^{-2}$ \cite{AlMo94}. $\Tc$ seems to
be determined more by the width of the narrow polaron band around the Fermi
level, which decreases exponentially with $g$. Thus one may expect that
$D/(k_{\rm B}\Tc)$ increases with increasing $g$, and thus with decreasing
$\Tc$. This is found experimentally, as discussed later.

Green and Edwards \cite{GrEd99} find that, in the DE model with $n=0.5$,
$\Tc$ only increases by $5\%$ when $S$ increases from $3/2$ to $\infty$. A
similar insensitivity to $S$ for $S\ge3/2$ is expected in the Holstein-DE
model. Hence, from equation~(\ref{eq:8}), with $A_E(\en)$ taken in the
$S\rightarrow\infty$ limit as discussed, we have $\delta=D/(k_{\rm B}\Tc
a^2)\propto(S+\frac{1}{2}n)^{-1}$. This is similar to the result
$\delta=\frac{1}{2}(s+1)^{-1}$ for the spin $s$ nearest neighbour simple
cubic Heisenberg model, with $\Tc$ calculated in mean field theory. Using
values of $\Tc$ accurate to within about $1\%$ \cite{RuWo58} we find an
improved Heisenberg value of $\delta=0.286$ for $s=3/2$, and $\delta=0.258$
for an interpolated $s=1.75$ which models the spin per site $S+\frac{1}{2}n$
in the present model. In the present calculations for $D/a^2$, together with
those of Green \cite{Gr01} for $\Tc$, we find, for $S=3/2$,
$\delta\approx0.24$ for $g/W=0.1$ and $\delta\approx0.29$ for $g/W=0.16$.
With $a=4$ {\AA} these values correspond to $D/(k_{\rm B}\Tc)\approx3.9$ and
$4.6$ {\AA}$^2$, respectively. The agreement of these values of $\delta$ with
the Heisenberg model suggests that for moderate $g/W$ the ferromagnetic
transition in the many-body CPA treatment of the Holstein-DE model is quite
Heisenberg-like. However, this may not be the case for larger $g/W$ where, as
discussed above, $\delta$ increases rapidly with $g/W$.

To compare our theoretical results with experiment we first note that the
simple spin-wave dispersion in equation~(\ref{eq:5}), which is of the
Heisenberg form, has been found to fit data on \chem{La_{0.7}Pb_{0.3}MnO_3}
($\Tc=355$ K) throughout the Brillouin zone \cite{PeAe96}. In this work the
low temperature stiffness constant $D=134$ meV{\AA}$^2$ so that $D/(k_{\rm
  B}\Tc)=4.4$ {\AA}$^2$. In LSMO, also with $x=0.3$, Martin \etal
\cite{MaEn96} find $\Tc=378$ K and $D\approx 188$ meV{\AA}$^2$ (at 27 K) so
that $D/(k_{\rm B}\Tc)=5.8$ {\AA}$^2$. \chem{Pr_{0.63}Sr_{0.37}MnO_3} seems
to behave similarly with $\Tc=301$ K, $D=165$ meV{\AA}$^2$, $D/(k_{\rm
  B}\Tc)=6.4$ {\AA}$^2$ \cite{FeDaHw98}. Although $D/(k_{\rm B}\Tc)$ in the
last two materials is larger than in a Heisenberg model, their spin dynamics
near $\Tc$ is quite conventional \cite{MaEn96,FeDaHw98}. However in some
systems with lower $\Tc$ this is not the case, and $D/(k_{\rm B}\Tc)$ is
considerably larger. Thus in LCMO, with $\Tc=250$ K, $D=170$ meV{\AA}$^2$,
$D/(k_{\rm B}\Tc)=7.9$ {\AA}$^2$, the ferromagnetic transition seems not to
be a standard second order one \cite{LyEr96}. Also in NSMO, with $\Tc=198$ K,
$D=165$ meV{\AA}$^2$, $D/(k_{\rm B}\Tc)=9.7$ {\AA}$^2$, the spin-wave
stiffness constant does not collapse to zero at $T=\Tc$ \cite{FeDaHw98}, just
as in LCMO \cite{LyEr96}. It is not clear whether such behaviour could occur
within the Holstein-DE model or whether inhomogeneity due to disorder is
important. However the larger values of $D/(k_{\rm B}\Tc)$ predicted by the
model for strong electron-phonon coupling suggests that something unusual is
going on. It is worth mentioning that $D$ may be underestimated by the
many-body CPA for an intermediate electron-phonon coupling such as $g/W=0.16$
appropriate to LCMO. This is because spurious incoherent scattering near the
Fermi level reduces the ``kinetic energy'' $K$, and hence $D$. On the other
hand equation~(\ref{eq:6}) itself gives an upper bound to $D$ which, in a
better approximation will certainly be reduced. For the DE model, in
approximations equivalent to the random phase approximation (RPA)
\cite{FuPom,Wa98,QuCeSi98}, this is achieved by an additional negative term
proportional to $\Delta^{-1}$ where $\Delta=JS$ is the Hartree-Fock exchange
splitting between up and down spin bands. In the presence of an onsite
Coulomb interaction $U$, $\Delta=JS+Un$ so that the negative term is
considerably reduced. RPA estimates of $D$ in a two-band model differ widely
\cite{QuCeSi98,Zh00}. For the one-band model with $J=\infty$ and $S=3/2$
Golosov \cite{Go00} has shown that $D$ is reduced to about half its
$S=\infty$ value, over a wide range of band-filling, when magnon-electron
scattering processes are considered. Further work on $D$ in the Holstein-DE
model is highly desirable.

\section{Conclusion}\label{sec:summary}

The many-body CPA treatment of the Holstein-DE model has been used to
investigate several spectral properties which may be compared with
experimental data on the manganites. We have been able to supply the theory
which was hinted at by Dessau \etal \cite{De98} in the discussion of their ARPES
measurements on the layered manganite \chem{La_{1.2}Sr_{1.8}Mn_2O_7} in the
low temperature ferromagnetic state. Broad spectral peaks lie either side of
a pseudogap at the Fermi level and the pseudogap contains polaron subbands
with exponentially small weight. One of these, at the Fermi level, is
responsible for the poor metallic behaviour. We therefore agree with
Alexandrov and Bratkovsky \cite{AlBrat99} that in this system, with unusually
strong electron-phonon coupling, small polarons exist in the ferromagnetic
state. However we find that small-polaron theory does not apply above or
below $\Tc$ in a pseudocubic manganite like \chem{Nd_{0.7}Sr_{0.3}MnO_3} with
intermediate coupling strength. In particular the small-polaron result that
the activation energy of the high-temperature dc conductivity is one-quarter
of the peak photon energy in optical conductivity $\si(\nu)$ is found not to
hold, in agreement with experiments on NSMO. The observed shift in spectral
weight of $\si(\nu)$ to lower energy on going into the ferromagnetic state is
found to occur, although it is somewhat suppressed by spurious incoherent
scattering at $T=0$ which is a defect of the theory. A rigorous upper bound
is derived for spin-wave energies at $T=0$ in the Holstein-DE model. It is
shown that the spin-wave stiffness constant $D$ decreases with increasing
electron-phonon coupling strength in a similar way to $\Tc$. However, for
strong coupling the ratio $D/(k_{\rm B}\Tc)$ increases quite rapidly with
increasing coupling strength, \ie with decreasing $\Tc$. This trend is found
experimentally.

\ack

We thank Lesley Cohen and Dietrich Meyer for helpful discussion, and Jim
Freericks for stimulating correspondence. One of us (MH) is grateful to the
International Association for the Exchange of Students for Technical
Experience (IAESTE), the Industriellenvereiningung Steiermark and Technical
University Graz for financial support during a visit to Imperial College.




\end{document}